# Topological spin dynamics in cubic FeGe near room temperature

Emrah Turgut[1], Matthew J. Stolt[2], Song Jin[2], Gregory D. Fuchs[1]

1. School of Applied and Engineering Physics, Cornell University, Ithaca, NY 14853, USA
2. Department of Chemistry, University of Wisconsin–Madison, WI 53706, USA

Understanding spin-wave dynamics in chiral magnets is a key step for the development of high-speed, spin-wave based spintronic devices that take advantage of chiral and topological spin textures for their operation. Here we present an experimental and theoretical study of spin-wave dynamics in a cubic B20 FeGe single crystal. Using the combination of waveguide microwave absorption spectroscopy (MAS), micromagnetic simulations, and analytical theory, we identify the resonance dynamics in all magnetic phases (field polarized, conical, helical, and skyrmion phases). Because the resonance frequencies of specific chiral spin textures are unique, quantitative agreement between our theoretical predictions and experimental findings for all resonance frequencies and spin wave modes enables us to unambiguously identify chiral magnetic phases and to demonstrate that MAS is a powerful tool to efficiently extract a magnetic phase diagram. These results provide a new tool to accelerate the integration of chiral magnetic materials into spintronic devices.

Identifying and understanding of spin-wave excitations in chiral and topological magnetic materials is a fundamental step towards their integration into spintronic devices. Moreover, some of these materials host magnetic skyrmions, in which spins form a topologically non-trivial excitation with an integer winding number.[1,2] Magnetic skyrmions can be driven to move by spin torques with an ultralow current threshold because there is weak or no pinning,[3,4] which makes them appealing for low-power memory,[5] magnetic logic,[6] and auto-oscillator[7] devices. In support of these potential applications, magnetic skyrmions have been investigated using several methods including Lorentz transmission electron microscopy (LTEM),[8–10] inelastic neutron scattering,[11,12] magnetic susceptibility,[13,14] and recently by microwave absorption spectroscopy.[15,16]

Microwave absorption spectroscopy (MAS), in particular, is a powerful tool for identifying and studying dynamical magnetic excitations in materials. In materials with several complex magnetic phases, such as chiral magnetic materials with a volume Dzyaloshinskii-Moriya interaction, it may be possible to use MAS to quickly establish a magnetic phase diagram because the spin texture in each magnetic phase has a unique resonance response. This specificity gives MAS an advantage as compared to electrical measurements such as the topological Hall effect,[17–22] which is more difficult to interpret because the signals are not unique to skyrmion phases or helical phases.[19–24]

In this letter, we present a study of spin-wave dynamics in a single crystal of B20 FeGe using MAS. Specifically, we measure the gigahertz frequency dynamic susceptibility of FeGe using a field-referenced lock-in technique, which significantly increases our sensitivity to magnetic phase boundaries as compared to conventional MAS.[15] We observe the resonance response in all the magnetic phases (field polarized, conical, helical and skyrmion phases) of FeGe, including a secondary skyrmion phase pocket previously observed by neutron scattering and magnetic susceptibility experiments in FeGe.[12,13] In addition, we adapted a recently developed analytical formalism for chiral spin dynamics and micromagnetic simulations to make definitive identification of the resonance frequencies for each phase. This combination of MAS and theoretical modeling enables us to uniquely identify the magnetic phase diagram of FeGe sensitively and efficiently.

We focus on FeGe because it has the highest critical temperature (280 K) among the noncentrosymmetric B20 compounds,[2,9,13,25] making the skyrmion phase accessible by cooling with a Peltier element. FeGe also has the closest lattice match to Si [111] substrate and it can be grown by magnetron sputtering for a scalable fabrication of relevant applications.[22,24,26] While FeGe has been the object of great interest in recent imaging, transport, and neutron scattering experiments,[9,10,25,27,28] the spin dynamics of its many magnetic phases are relatively unexplored.

The single crystal of B20 FeGe was grown by chemical vapor transport as described by Richardson.[29] We determined the structure and crystallographic orientation of a millimeter-sized, pyramid-shaped B20 FeGe crystal using Laue X-ray diffraction prior to MAS experiments (see Fig. S1). To perform MAS, we place the crystal inside an environmentally-controlled sample box in which it sits directly on a broadband coplanar waveguide (CPW, Fig. 1a). We apply both a

D.C. magnetic field ($H_{DC}$) and a weak a.c. magnetic field with amplitude $\Delta H_{ac}$ = 6 Oe at frequency $2\pi\omega_s$ = 503 Hz. The sample temperature is controlled using a thermoelectric cooling element. To excite magnetic dynamics, we apply radio-frequency (RF) power to the CPW, which generates an RF magnetic field ($H_{RF}$) just above the coplanar waveguide. Our set-up enables measurement frequencies in the range of 0.1 – 18 GHz; additional details can be found in Ref.[30]. We align the [110] crystal axis along the DC magnetic field. We also performed the experiment when the [110] axis and the D.C. field were perpendicular, but obtained the same results.

To obtain the crystal's gigahertz dynamic susceptibility we measure the transmitted microwave power using an RF diode. We amplify its output and pass the resulting signal to a lock-in amplifier that demodulates with respect to $\Delta H_{ac}$ (Fig. 1b). This corresponds to the field derivative of transmitted RF power, (dP/dH), in the in-phase (X) and the out-of-phase (Y) components of the lock-in amplifier, that rejects non-magnetic contributions. As illustrated in Fig. 1c, we scan the DC magnetic field, the frequency of $H_{RF}$, and the temperature of the sample to map the resonances that exist in the magnetic phases of bulk FeGe, including field polarized, helical, conical, skyrmion, and paramagnetic phases (Fig. 1d).[9,12,13,31]

We first study the field-polarized phase of the FeGe single crystal, which appears at large magnetic fields both above and below the critical temperature $T_c$ = 280 K, which is defined by the formation of the helical phase under zero-field. To isolate the field-polarized phase, we acquired MAS data at magnetic fields above 1500 Oe and frequencies between 6–11 GHz, thus rejecting the conical phase resonances. Because we measure dP/dH, we fit the results to a derivative of Lorentzian lineshape (Fig. 1e at 264 K) to extract the resonance fields ($H_0$) and the linewidths ($\Delta H$) [See SM]. Next, to find intrinsic and extrinsic linewidths, we use $\Delta H = \frac{4\pi\alpha f_0}{\gamma\mu_0} + \Delta H_0$, where $\gamma$ is electron's gyromagnetic ratio, $\mu_0$ is the vacuum magnetic susceptibility, $\alpha$ is the Gilbert damping factor, and $\Delta H_0$ is inhomogeneous broadening.[32] Determining the intrinsic damping is relevant for applications, because it influences power dissipation, spin torque critical currents, and the velocity of chiral domain walls.[33] Therefore, we also study the temperature dependence of $\alpha$ and $\Delta H_0$ at 256–286 K temperature range (Fig. 1f). While $\alpha$ increases monotonically up to $T_c$, it drops substantially above $T_c$. In contrast, $\Delta H_0$ has the opposite behavior.

We note that $\alpha$ has complicated behavior above the $T_c$, which may complicate the interpretation of helimagnetic behavior near the $T_c$. For example, Zhang et al.[34] recently reported an ultrasmall damping (α = 0.0036±0.0003) in FeGe thin films on MgO substrate at 310 K, which may significantly differ at lower temperatures. Moreover, our measurement of $\alpha$ in FeGe single crystal is 0.066±0.003 at 264 K, which is three times larger than the value we recently reported in sputtered FeGe films on Si [111] (α = 0.021±0.005) at 263 K,[26] and which could be influenced by the lower $T_c$ in thin films with substrate induced strain and anisotropy.

Next, we investigate the spin dynamics in the helical and conical phases. First, we model the spin dynamics with micromagnetic simulations using Mumax3.[35] We use the material parameters for T = 274 K to rigorously replicate the experiment [See SM]. After initializing the magnetic state, we relax the system into its ground state. In Fig. 2a, we show that the corresponding equilibrium magnitude of the z-component (both the magnetic field direction and the direction of the helical vector in z) of the magnetization ($M_z$) increases linearly with the magnetic field and reaches the saturation at 850 Oe. This value is consistent with the saturation magnetic field $H_d$ = 840 Oe that we measured in the experiment (Fig. 2d). Next, we computationally apply a 5 Oe magnetic pulse, which excites spin precession, and monitor the spatially and temporally resolved magnetization dynamics (the ring-down method).[36] By taking the discrete Fourier transform at each pixel, we find the resonance frequencies and corresponding mode profiles as a function of magnetic field. From these data, we construct a power spectral density plot as shown in Fig. 2b. We observe two chiral modes, i.e. $Q_-$ and $Q_+$, which are clockwise (CW) and counter-clockwise (CCW) precession of a spin around its equilibrium orientation.[15] Moreover, in Fig. 2c, we plot the result of a separate calculation of the resonance frequencies for these two chiral modes as a function of the ratio of magnetic field with respect to the saturation field ($H_{c2} \equiv H_d$) using the analytic model derived in Ref.[15] for an isotropic sample. The analytical calculation is in close agreement with the micromagnetic calculation.

Next, we perform MAS in the range of 0.1 - 6.3 GHz and −1200 - 1200 Oe. We note that the $Q_-$ mode has a strong response at low field, but decays as the field approaches the saturation field. The $Q_+$ mode behaves oppositely. This agrees with the analytic calculation of the mode weights in Ref.[37]. Interestingly, while we observe conical and field polarized phase resonances strongly in the in-phase component (X) of the signal (Fig. 2d), the out-of-phase component (Y) of the signal (Fig. 2e) most strongly reflects the helical to conical phase boundary.

We explain the difference between the X and Y component response by analyzing our detection scheme. For the absorbed microwave power P(H) as a function of H, the X and Y of the $\Delta H_{ac}$ referenced lock-in components are proportional to $X = \int_0^{\frac{2\pi}{\omega_s}} P(H) \cos \omega_s t \, dt$ and $Y = \int_0^{\frac{2\pi}{\omega_s}} P(H) \sin \omega_s t \, dt$. We expand P(H) as Taylor series,

$$P(H) = P(H_0) + \frac{dP}{dH}(H - H_0) + \frac{d^2P}{dH^2}\frac{(H-H_0)^2}{2} + \cdots \quad (1)$$

where $H - H_0 = \Delta H_{ac} \cos \omega_s t$. If we assume P(H) is a well-defined and continuous function for all H, the X integral will be non-zero for odd orders of $H - H_0$, while the Y integral vanishes for all the orders. This is true for the field-polarized phases and their boundaries when the transition is gradual. It is also consistent with their strong response in only the X component of our signal. On the other hand, for the helical phase with multiple q vector (and also for the skyrmion phase in the next section), microwave absorption P(H) is not a continuous function at magnetic phase boundaries because the boundaries are abrupt. Thus, neither the X or the Y components of the lock-in vanish at such a "kink" in P(H). Although the extra contribution at the phase boundary is present in X, it is small compared to the direct resonant absorption signal. The Y component, however, is nearly zero except at abrupt phase boundaries, which makes abrupt magnetic phase boundaries stand out. Teasing apart phase boundaries from direct magnetic resonance is more challenging using the vector network analyzer (VNA) technique that have been used previously,[15] because it probes P(H) directly; thus the two methods are complimentary. For a direct comparison, see the supporting material in which we present VNA measurements of *P(H)* side-by-side with lock-in measurements of *dP/dH*.

Finally, we study spin dynamics in the magnetic skyrmion phase with micromagnetic simulations (Fig. 3) and MAS experiments (Fig. 4). In our simulations, we use FeGe material parameters at 277 K and employ the ringdown method again. We find the equilibrium magnetic configuration, a hexagonal skyrmion lattice, by relaxing the system into its ground state under a 300 Oe out-of-plane magnetic field (Fig. 3a). The skyrmion size is 70 nm, which matches the value measured using LTEM,[9] which verifies the D and J terms in our model. We also added a minuscule easy axis uniaxial anisotropy of $K_u$ = 200 J/m$^3$, which helps to numerically stabilize the skyrmion phase against the conical phase that is energetically favored with zero $K_u$.[13,31,38] In our Fourier analysis, to exclude boundary effects, we focus on the central skyrmion, whose 3-dimensional spin configuration is also shown to confirm its topological state (Fig. 3a).

To gain insight into the skyrmion modes, we calculated the spatially averaged spectra of the dynamics (Fig. 3b). Because of the thickness profile of the sample, we observe very sharp resonances corresponding to thickness modes. These are an artifact of limited simulation volume, and they are marked with pink dots in Fig. 3b. Additionally, we find three resonances at 1.925, 2.55, and 3.55 GHz frequencies, whose spin-wave modes are in the plane and uniform through the thickness. The spatial profile of the 1.925 GHz resonance is shown in Fig. 3c1-c3 by taking Fourier transform of individual $m_x$, $m_y$, and $m_z$ magnetizations. Using these mode images directly, we identify the spin-wave mode as either CW or CCW, i.e. the center of skyrmion is large in $m_x$ and $m_y$, while $m_z$ is large in the intermediate regions. To find the exact nature of this spin-wave mode, we drive the system with an $\{H_x, H_y, H_z\} = \{0, 5\sin(2\pi \times t \times 1.925 \text{ GHz}), 0\}$ Oe magnetic field with a realistic damping coefficient $\alpha = 0.066$ and record the magnetization eight times per period for a total 36 periods. In so doing, we eliminate all other resonances but the 1.925 GHz one, and more closely model the experiment. Then, we plot the deviation of magnetization from its equilibrium state at each time step for a full period in Fig. 3d. We observe that the resonance at 1.925 GHz is the CCW mode. This conclusion is consistent with analytic calculations in Refs.[15,37], in which the CCW mode has a significantly larger spectral weight.

We also tried to computationally drive the system in a similar way at 2.55 and 3.55 GHz, but the resulting dynamics always lead to a CCW mode due to the weak spectral weight of these modes. However, using a previously reported analytic model[15,37] and numerical simulations,[39] the CW mode is expected to have resonances around 3.56 GHz (~1.85 times of CCW mode frequency), which agrees well with the third resonance at 3.55 GHz. We could not identify the 2.55 GHz resonance, which may be a micromagnetic artifact or a mix of CCW and CW modes.

Next, we perform MAS in the temperature range of skyrmion formation, 279 – 276 K, and plot the results as a function of frequency and magnetic field (Fig. 4). As before, the *Y* component is sensitive to abrupt magnetic phase boundaries with the helical or skyrmion phases (Fig. 4a-d), whereas the *X* component is most sensitive to the field polarized and conical phases (Fig. 4e-h). In addition, at 276 and 277 K the *X* component has some conical resonances between the helical and skyrmion phases and there are multiple competing phases at narrow field ranges as previously found in $Cu_2OSeO_3$.[16]

We note that there are two dominant resonances within the skyrmion phases at 1.8 and 3 GHz at 278 K, corresponding to the CCW and CW modes, respectively. Because we drive the system with an in-plane RF field, we do not expect to observe breathing modes, which can be excited only by an out-of-plane $H_{RF}$ field.[15,16,37] In addition, we observe an additional skyrmion pocket at higher fields; an observation that is different from previous spin dynamics studies in other bulk B20 materials.[15,16] For example, the MAS at 279 K reveals only two skyrmion resonances (CCW and CW in Fig. 4a), whereas at lower temperatures a total of four skyrmion resonances appear between the conical and helical phases (Fig. 4b–d). Moreover, at 276 K, the original skyrmion resonance becomes weak, but the high field skyrmion resonance is more evident. These multiple skyrmion phases were previously found in susceptibility, specific heat, and neutron scattering experiments in bulk FeGe and named $A_1$ (low field) and $A_2$ (high field) phases, respectively.[12,13,31] The origin of the $A_1$ and $A_2$ phases emerges from the stabilization of skyrmions by thermodynamic fluctuations at the vicinity of $T_c$. It was found that the interaction between skyrmions differs at higher magnetic fields and this results in the different susceptibilities, specific heats, and neutron scattering signatures.[12,31,38,40,41] We find that this effect is also accessible by MAS.

We plot the magnetic phase diagram of the FeGe single crystal determined by MAS in Fig. 5 as a function of magnetic field and temperature. Our phase diagram is identical to the one determined by neutron scattering,[12] however, using magnetic susceptibility Wilhelm *et al.* observed four skyrmion phase pockets; the extra two were labeled $A_0$ and $A_3$ and they were located just below (in field) the $A_1$ phase and to the left of (in temperature) the $A_1$ phase.[13] In a separate specific heat study,[31] only the $A_1$ and $A_2$ phases were found, consistent with our study.

In summary, we studied spin dynamics in a non-centrosymmetric B20 FeGe single crystal using microwave absorption spectroscopy. We find that phase-sensitive measurement of differential microwave absorption is an efficient method to establish chiral magnetic phase boundaries, including helical and skyrmion phases. Additionally, the quantitative correspondence between experimentally measured resonance frequencies and theoretical calculations enable unambiguous phase identification. Our findings are relevant to the development of skyrmionic logic and storage devices, spin-torque oscillators, and nonreciprocal magnetic devices because a quantitative understanding of spin-waves in chiral magnets will be crucial to creating active spintronic devices with topological phases.

**Supplementary Material**

See supplementary material for X-ray characterization of the FeGe single crystal sample, calculations of the temperature dependent material parameters, how some of these parameters were used in the micromagnetic simulations, additional VNA MAS measurements, and spin-wave modes for 2.55 and 3.55 GHz frequencies in the skyrmion phase.

**Acknowledgement**

The microwave absorption spectroscopy experiment, micromagnetic simulation, and data analysis were supported by the DOE Office of Science (Grant # DE-SC0012245) and the single crystal FeGe growth was supported by the NSF (Grant # ECCS-1609585). We also acknowledge use of facilities of the Cornell Center for Materials Research (CCMR), an NSF MRSEC (Grant # DMR-1120296). We further acknowledge facility use at the Cornell Nanoscale Science and Technology Facility (Grant # ECCS-1542081), a node of the NSF-supported National Nanotechnology Coordinated Infrastructure. M.J.S. also acknowledges support from the NSF Graduate Research Fellowship Program (Grant # DGE-1256259).

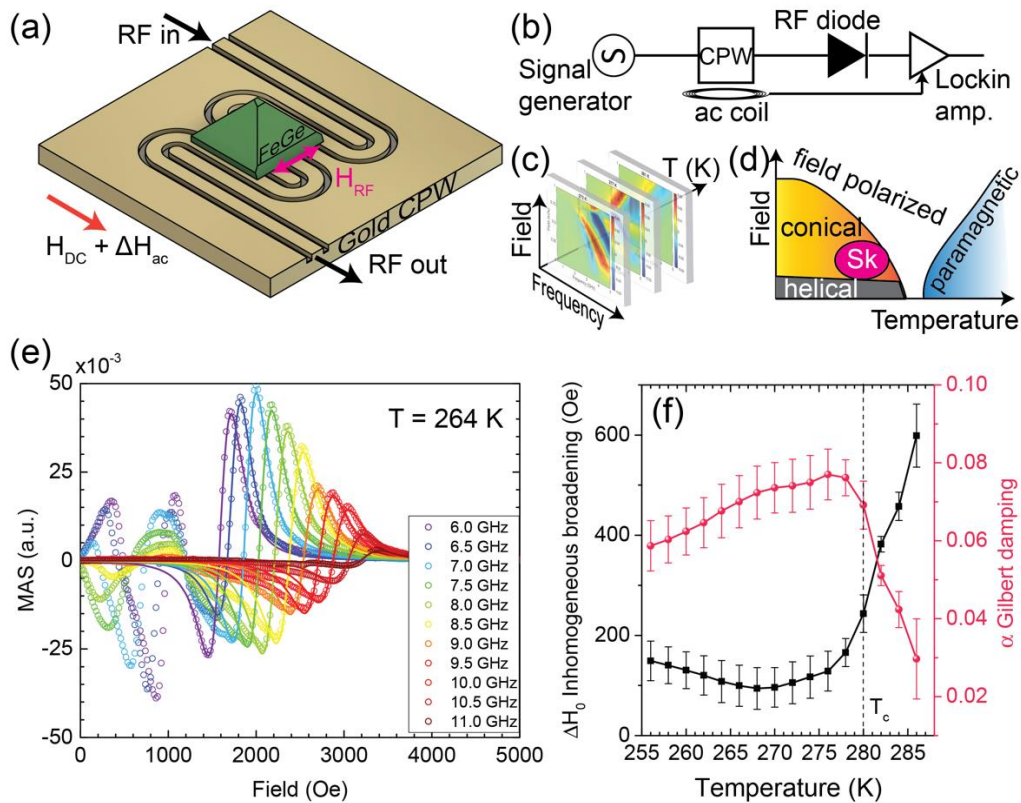

Figure 1. (a) A mm size, single crystal, pyramid-like FeGe sample is placed above a broadband coplanar waveguide that is cooled with a Peltier element and under a DC and ac magnetic field. (b) Microwave absorption spectroscopy (MAS): microwave power absorption is measured using an RF diode and a lock-in amplifier that is referenced to the ac magnetic field. (c) MAS data are collected by varying the DC magnetic field, the RF frequency and the temperature of the sample in order to map multipart-phase diagram of FeGe (d). (e)The uniform (Kittel) mode of the field polarized phase at T = 264 K and above the saturation magnetic fields. The derivative fit to the power spectra are shown by the solid curves. (f) Temperature dependence of $\alpha$–the Gilbert damping and $\Delta H_0$–the inhomogeneous broadening.

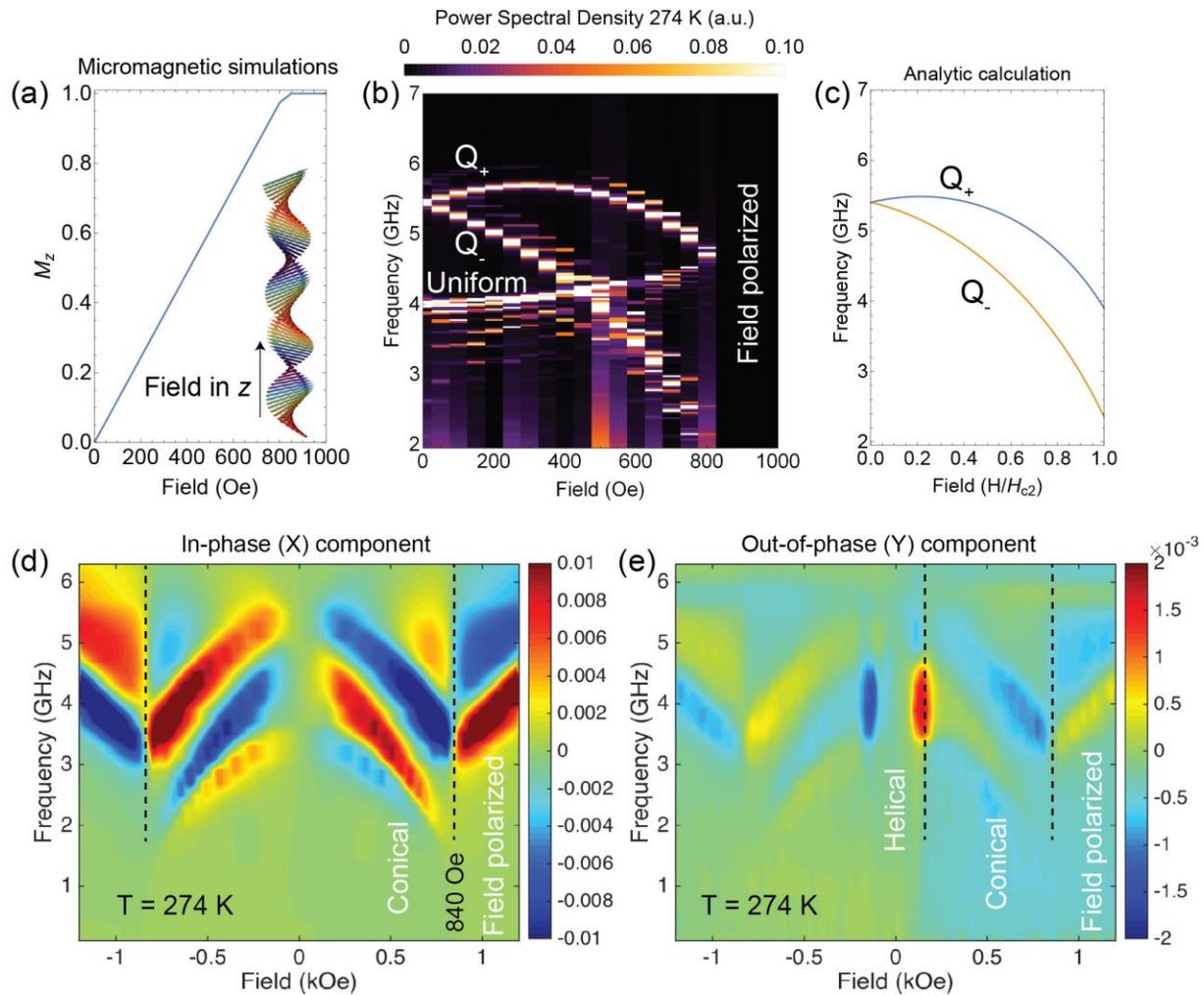

**Figure 2.** Micromagnetic simulations, analytic calculations, and experimental spectra of the helical and conical phases at 274 K. (a) In micromagnetic simulation, the magnetic component along the DC field is linear with the applied field. The spin configuration is shown by an inset in which spins curl around the *z*-axis. (b) Power spectral density of the helical and conical phases is obtained by ring-down micromagnetic simulations. Uniform, $Q_+$, and $Q_-$ modes are the dominant modes. (c) We also calculate and plot the resonance frequencies of $Q_+$ and $Q_-$ modes using theory from Ref.[15] as described in the text. (d) In-phase (X) and (e) out-of-phase (Y) lock-in components of the RF diode signal.

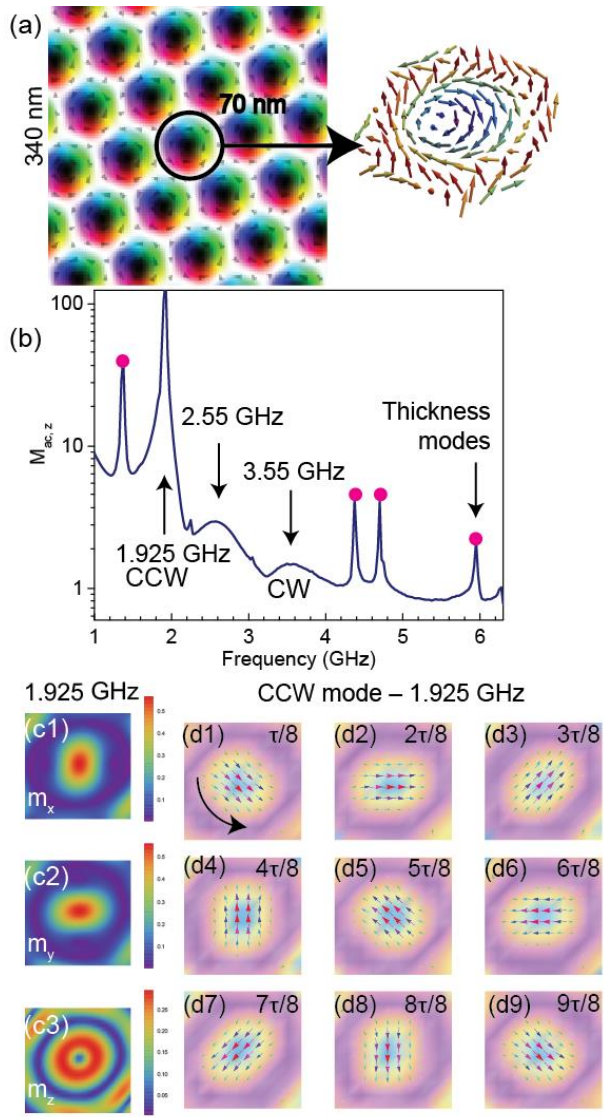

Figure 3. Micromagnetic simulations of the skyrmion phase using materials parameters at 277 K as discussed in the text. (a) Initial relaxed state under a 300 Oe magnetic field perpendicular to the plane. The individual skyrmion size is 70 nm. (b) Dynamical spin spectrum. Black arrows mark the CCW and CW modes. The sharp resonances marked with pink dots are due to simulation artifacts of the confined geometry. (c) Spatially resolved Fourier transform of the magnetization dynamics of x, y, and z components at 1.925 GHz. (d) The counter-clockwise rotation of the skyrmion under a sinusoidal magnetic field with an amplitude of 5 Oe and a frequency of 1.925 GHz. From (d1) to (d9), counter-clockwise rotation is demonstrated.

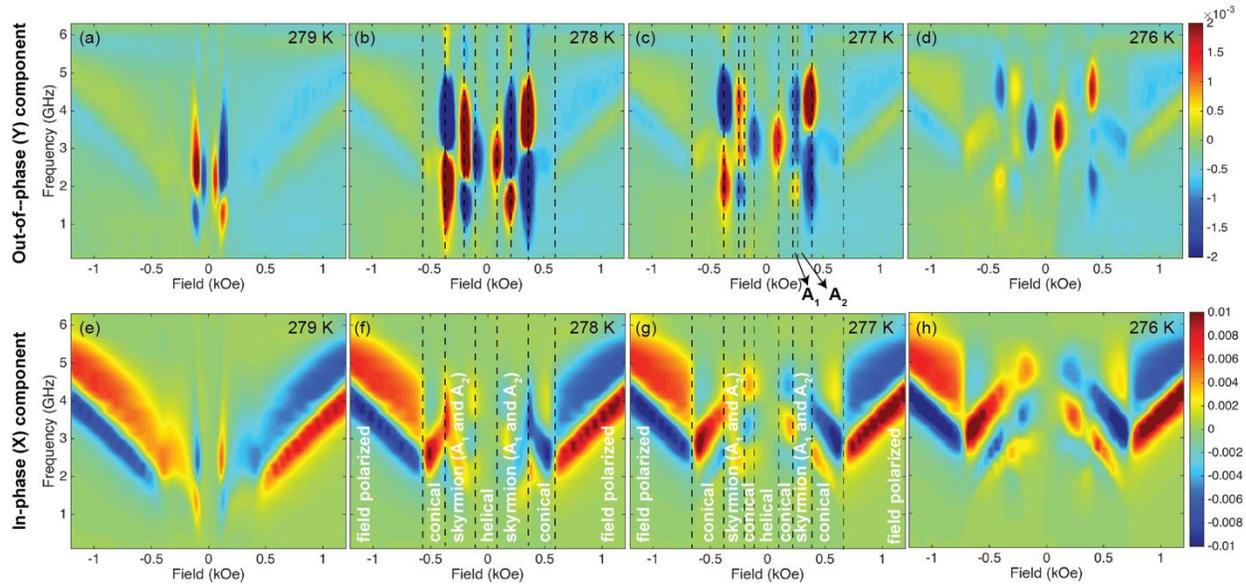

Figure 4. Experimental spectra of FeGe at temperatures close to the critical temperature of 280 K, which show helical, skyrmion, conical, and field polarized magnetic phases. The first row (a–d) shows the out-of-phase (Y) component, which is sensitive to the helical and skyrmion phases. The second row (e-h) shows the in-phase (X) component, which is more sensitive to the field polarized and conical phases. We identify the phase boundaries by dashed lines in (f) at 278 K by also using (b).

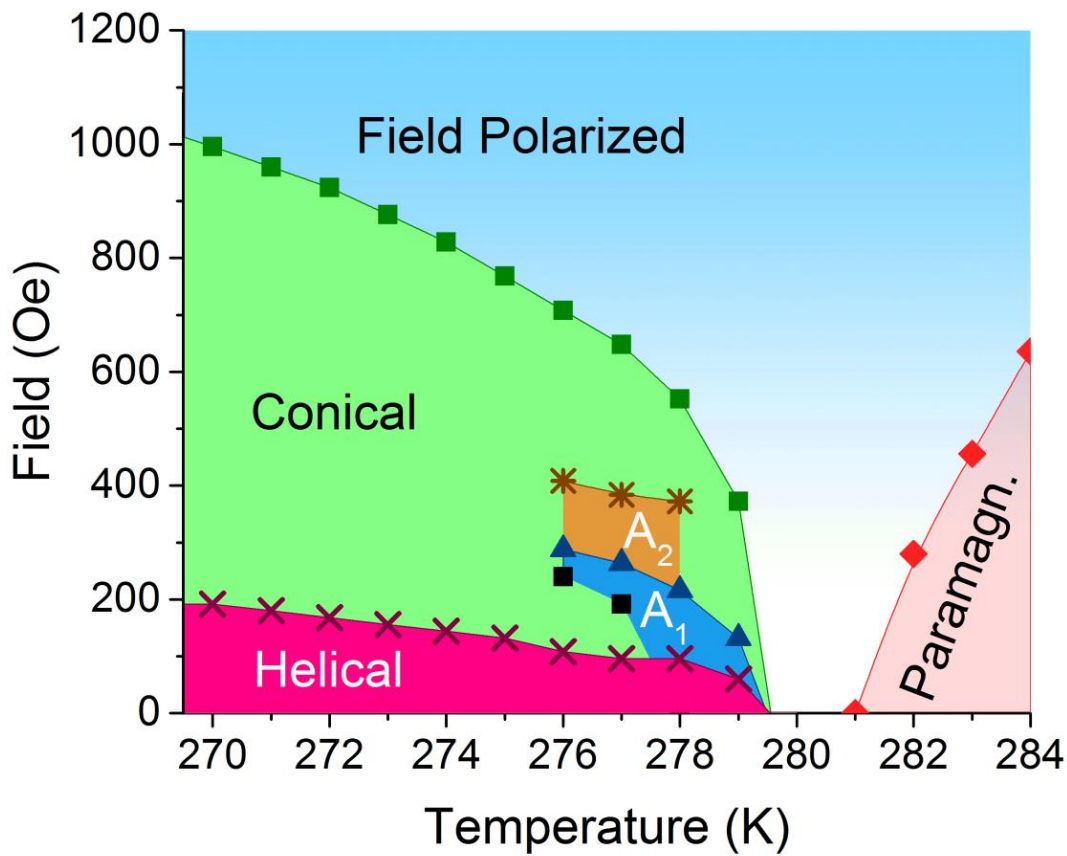

**Figure 5. Magnetic phase diagram of FeGe single crystal as a function of DC magnetic field and temperature determined using MAS.**

# Supplementary Materials: Topological spin dynamics in cubic FeGe near room temperature

Emrah Turgut[1], Matthew J. Stolt[2], Song Jin[2], Gregory D. Fuchs[1]

1. School of Applied and Engineering Physics, Cornell University, Ithaca, NY 14853, USA
2. Department of Chemistry, University of Wisconsin–Madison, WI 53706, USA

## 1. X-ray characterization of the B20 FeGe single crystal

We measured the crystallographic orientation of the FeGe single crystal sample with a Laue X-ray diffractometer. First, we collect the diffraction points and find the unique solution of these points, which are marked by red in Fig. S1. The corresponding alignment of the FeGe single crystal is shown next to its stereographic projection.

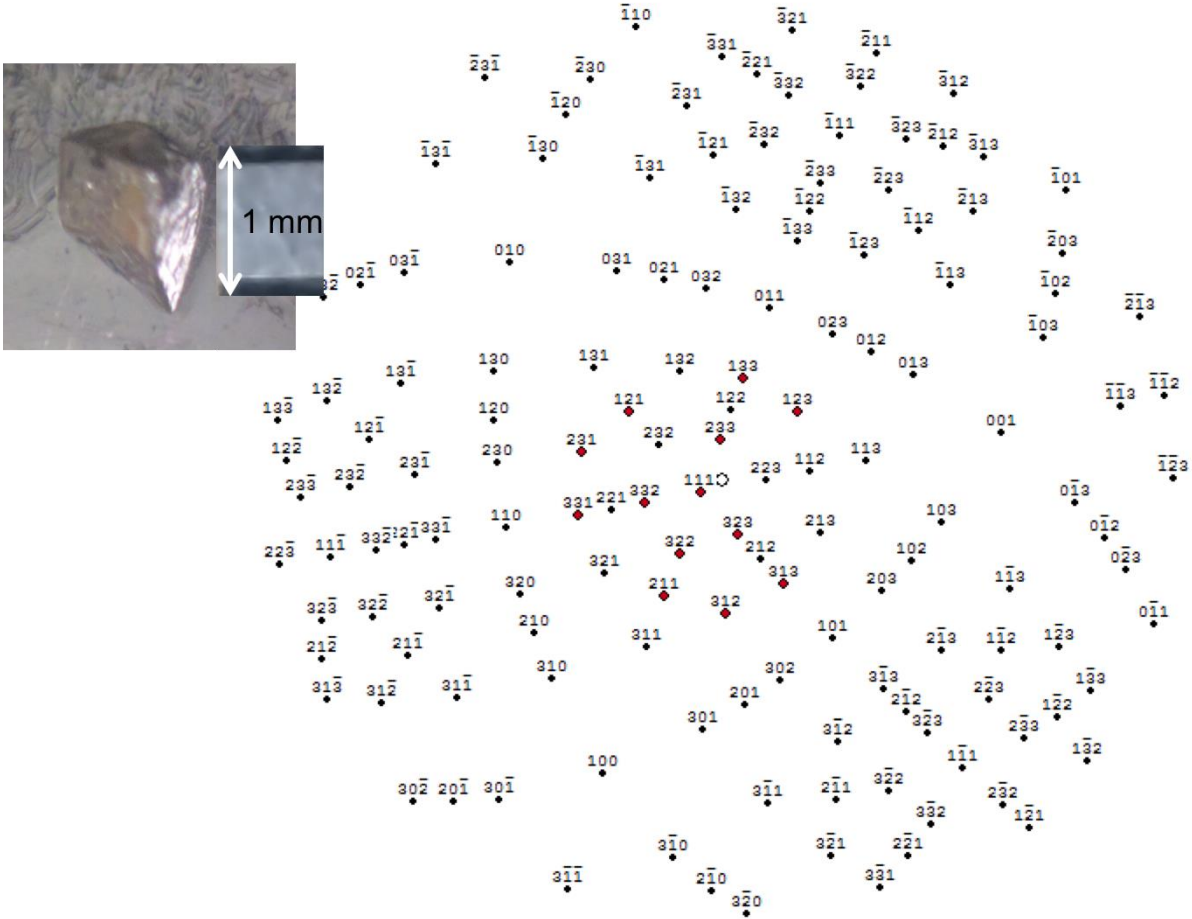

**Figure S1.** Stereographic projection of the crystallographic orientations of the FeGe single crystal and the actual image of the crystal at this orientation. The red dots are measured diffraction spots by a Laue diffractometer and shown axis are the unique solution for $P2_13$ cubic crystal structure.

## 2. Linewidth extraction

We use the symmetric and the asymmetric combination of the derivative of Lorentzian line shape to fit our field polarized ferromagnetic resonance curves, i.e.:

$$f(H_0, \Delta H) = a \frac{(H-H_0)\left(\frac{\Delta H}{2}\right)^2}{\left[\left(\frac{\Delta H}{2}\right)^2 + (H-H_0)^2\right]^2} + b \frac{\left(\frac{\Delta H}{2}\right)^3}{\left[\left(\frac{\Delta H}{2}\right)^2 + (H-H_0)^2\right]^2}, \quad (S1)$$

where $\Delta H$ is the linewidth, $H_0$ is the resonance fields, $a$ and $b$ are arbitrary coefficients. To find intrinsic and extrinsic linewidths, we use $\Delta H = \frac{4\pi\alpha f_0}{\gamma\mu_0} + \Delta H_0$, where $\gamma$ is the electron gyromagnetic ratio, $\mu_0$ is the vacuum magnetic susceptibility, $\alpha$ is the Gilbert damping factor, and $\Delta H_0$ is inhomogeneous broadening.

## 3. Extracting $M_s$ and $H_d$ from previous studies

The saturation magnetization ($M_s$) and magnetic field ($H_d$) strongly depend on the sample temperature, thus using the right material parameters is extremely important for accurate micromagnetic and analytic calculations. We assume that the magnetization and saturation magnetic field depends on the temperature as:

$$M_s = M_{s0}\left(1 - \frac{T}{T_c}\right)^{\beta_1}, \text{ and } H_d = H_{d0}\left(1 - \frac{T}{T_c}\right)^{\beta_2}, \quad (S2)$$

where $T_c$ is the critical temperature, $M_{s0}$ and $H_{d0}$ are phenomenological constants, and {$\beta_1$, $\beta_2$} are critical exponents[1,2]. Then, we extract the magnetization data from Ref.[2] and the saturation field from Ref.[1] and obtain the following relations:

$$M_s = 468607 \, A/m \left(1 - \frac{T}{T_c}\right)^{0.330}, \text{ and } H_d = 2974 \, Oe \left(1 - \frac{T}{T_c}\right)^{0.329} \quad (S3)$$

Because two References [1,2] found different values of $T_c$, which could be a result of different experimental configurations, we did not use their values. Instead, we used our microwave absorption spectroscopy measurements, in which the helical and conical resonances disappear at 280 K.

Lastly, we plug 280 K as $T_c$ in the equations above and find the material parameters, which are shown in Table S1. We also calculated the critical frequency, $f_c$, and dimensionless susceptibility, $\chi \equiv 4\pi M_s / H_d$.

**Table S1.** Calculated saturation magnetization $M_s$, saturation field $H_d$, exchange stiffness, critical frequency $f_c$, and the susceptibility $\chi$.

| T (K) | $M_s$ (A/m) | $H_d$ (Oe) | $A_{ex}$ (J/m) | $f_c$ (GHz) | $\chi$ |
|---|---|---|---|---|---|
| 279 | 72986.1 | 466.1537 | 2.11E-13 | 1.304893 | 1.967528 |
| 278 | 91744.5 | 585.5148 | 3.33E-13 | 1.639017 | 1.969029 |

| 277 | 104879.4 | 669.0434 | 4.35E-13 | 1.872837 | 1.969907 |
| 276 | 115324.1 | 735.439  | 5.26E-13 | 2.058696 | 1.970531 |
| 275 | 124136.7 | 791.4443 | 6.10E-13 | 2.215471 | 1.971014 |
| 274 | 131834.8 | 840.3555 | 6.88E-13 | 2.352387 | 1.97141  |
| 273 | 138714.7 | 884.0602 | 7.61E-13 | 2.474728 | 1.971744 |
| 272 | 144963.9 | 923.752  | 8.31E-13 | 2.585836 | 1.972034 |
| 271 | 150709.3 | 960.2393 | 8.98E-13 | 2.687974 | 1.972289 |
| 270 | 156041.5 | 994.0978 | 9.63E-13 | 2.782754 | 1.972518 |
| 269 | 161027.4 | 1025.754 | 1.03E-12 | 2.871368 | 1.972724 |

4. **Parameters used in micromagnetic simulations**

For the conical and helical phases at 274 K:

```
Ku1=0
Msat=132e3
exccons:=6.88e-13
Aex=exccons
DD:=4*pi*exccons/70e-9
Dbulk=DD
anisU=vector(0,0,1)
alpha = 0.002
```

For the skyrmion phase at 277 K:
The volume of the simulation is 340x340x20 nm³ by a total 170x170x10 cells, which consists of approximately 19 skyrmions.
```
Ku1=200
Msat=104.9e3
exccons:=4.35e-13
Aex=exccons
DD:=4*pi*exccons/70e-9
Dbulk=DD
anisU=vector(0,0,1)
alpha = 0.002
B := 0.03
```

5. **Spin-wave modes of 2.55 and 3.55 GHz frequencies in skyrmion phase**

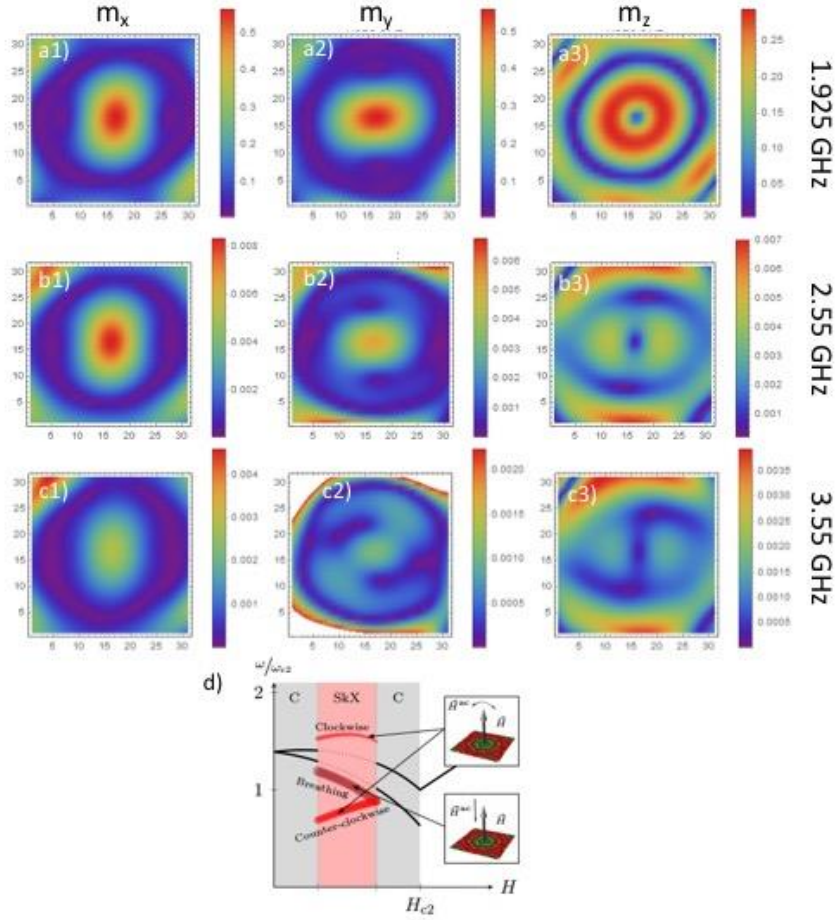

**Figure S2.** Spin-wave modes of $m_x$ (first column), $m_y$ (second column), and $m_z$ (third column) at 1.925 GHz (first row), 2.55 GHz (second row), and 3.55 GHz (third row) frequencies. (d) Analytic calculation of skyrmion resonances taken from Schwarze et al.[3] Because we have only in-plane $H_{rf}$ excitation field, we expect to have only CCW and CW modes. We found $f_c$ = 1.87 GHz using the saturation field and susceptibility at 277 K (Table S1). We expect 1.95 GHz to be CCW and 3.55 GHz to be CW.

6. **Comparison of high-frequency susceptibility and VNA measurements**

In this section, we compare our experimental method based on locking into an RF diode with the VNA method used in Ref.[3] in Figure S3 for the conical phase and in Figure S4 for the skyrmion phase.

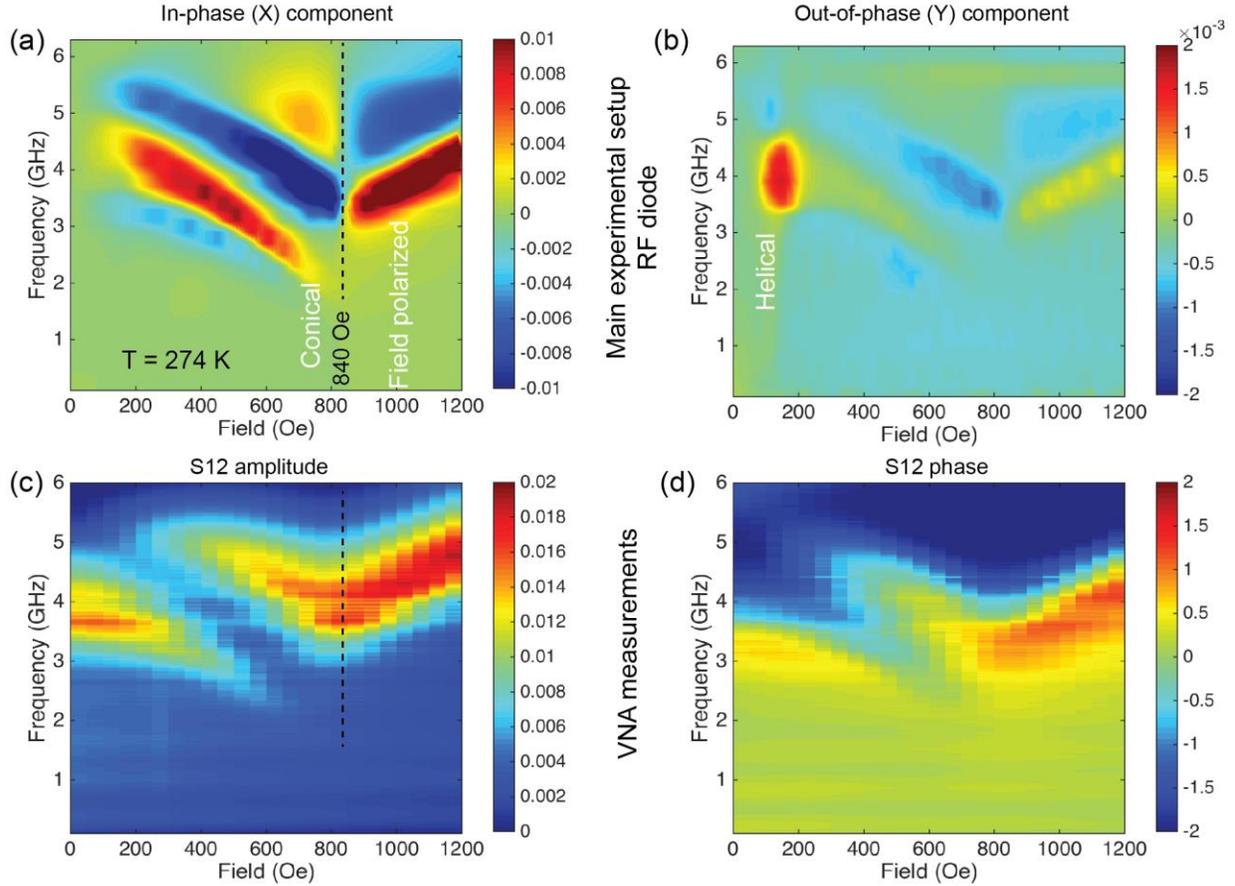

**Figure S3.** Comparison of the RF diode lock-in approach and VNA measurements for the conical phase. While the VNA shows the conical and field polarized phases clearly, the multi-domain helical phase resonance and phase boundary is better resolved by our RF diode lock-in method.

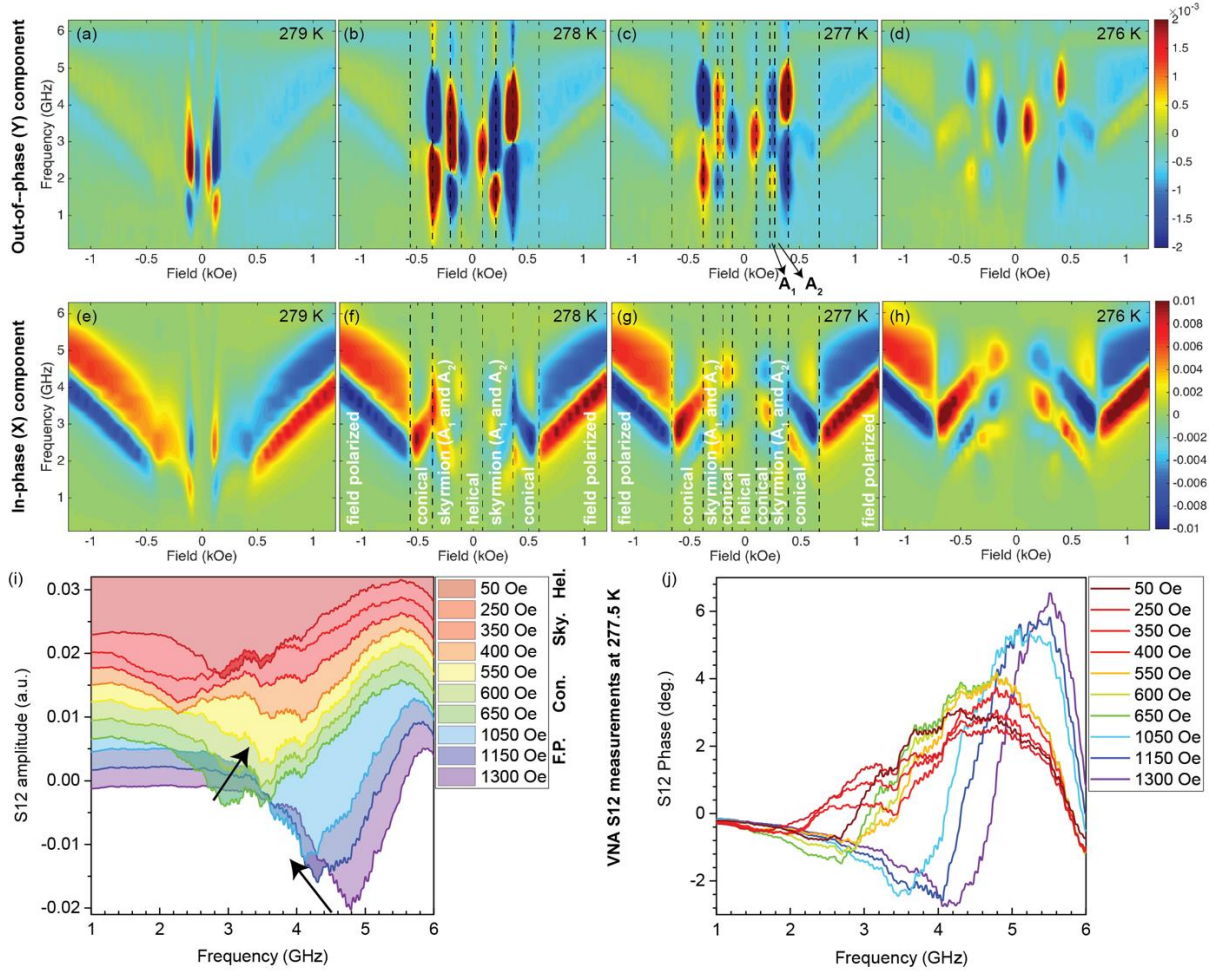

**Figure S4. Comparison of the RF diode lock-in and VNA measurements for the skyrmion phase. While the VNA resolves only the joined skyrmion phases, the second skyrmion phase pocket was only observed by RF lock-in diode method. (a)-(h) are the same as in the main text Fig. 4. (i) and (j) are the amplitude and the phase of the S12 parameter measured using a VNA.**